\documentclass[aps,preprint]{revtex4}%
\usepackage{amsfonts}
\usepackage{amsmath}
\usepackage{amssymb}
\usepackage{subfigure}
\usepackage{graphicx}
\usepackage{array}%
\begin{document}
\preprint{CTP-SCU/2019018}
\title{Thermodynamic Geometry of AdS Black Holes and Black Holes in a Cavity}
\author{Peng Wang}
\email{pengw@scu.edu.cn}
\author{Houwen Wu}
\email{iverwu@scu.edu.cn}
\author{Haitang Yang}
\email{hyanga@scu.edu.cn}
\affiliation{Center for Theoretical Physics, College of Physics, Sichuan University,
Chengdu, 610064, China}

\begin{abstract}
The thermodynamic geometry has been proved to be quite useful in understanding
the microscopic structure of black holes. We investigate the phase structure,
thermodynamic geometry and critical behavior of a Reissner-Nordstrom-AdS black
hole and a Reissner-Nordstrom black hole in a cavity, which can reach
equilibrium in a canonical ensemble. Although the phase structure and critical
behavior of both cases show striking resemblance, we find that there exist
significant differences between the thermodynamic geometry of these two cases.
Our results imply that there may be a connection between the black hole
microstates and its boundary condition.

\end{abstract}
\keywords{}\maketitle
\tableofcontents

\bigskip



\section{Introduction}

The study of black hole thermodynamics has been playing an increasingly
prominent role in our understanding of the interdisciplinary area of general
relativity, quantum mechanics, information theory and statistical physics. In
the pioneering work
\cite{IN-Hawking:1974sw,IN-Bekenstein:1972tm,IN-Bekenstein:1973ur}, black
holes were found to possess thermodynamic properties such as entropy and
temperature. Later, the Hawking-Page phase transition (i.e., a phase
transition between the thermal anti-de Sitter (AdS) space and a black hole)
was discovered in Schwarzschild-AdS black holes \cite{IN-Hawking:1982dh}.
Unlike Schwarzschild black holes, Schwarzschild-AdS black holes can be
thermally stable since the AdS boundary acts as a reflecting wall for the
Hawking radiation. With the advent of the AdS/CFT correspondence
\cite{IN-Maldacena:1997re,IN-Gubser:1998bc,IN-Witten:1998qj}, there has been
much interest in studying the thermodynamics and phase structure of various
AdS black holes
\cite{IN-Witten:1998zw,IN-Chamblin:1999tk,IN-Chamblin:1999hg,IN-Caldarelli:1999xj,IN-Cai:2001dz,IN-Kubiznak:2012wp,IN-Wang:2018xdz,IN-Wang:2019jzz}%
. Specifically, it was found that Reissner-Nordstrom-AdS (RN-AdS) black holes
exhibit a van der Waals-like phase transition (i.e., a phase transition
consisting of a first-order phase transition terminating at a second-order
critical point) in a canonical ensemble
\cite{IN-Chamblin:1999tk,IN-Chamblin:1999hg} and a Hawking-Page-like phase
transition in a grand canonical ensemble \cite{IN-Peca:1998cs}.

Alternatively, one can make asymptotically flat black holes thermally stable
by placing them inside a cavity, on the wall of which the metric is fixed.
York first showed that Schwarzschild black holes in cavity can be thermally
stable and have quite similar phase structure and transition to these of
Schwarzschild-AdS black holes \cite{IN-York:1986it}. For Reissner-Nordstrom
(RN) black holes in a cavity, the thermodynamics and phase structure have been
studied in a grand canonical ensemble \cite{IN-Braden:1990hw} and a canonical
ensemble \cite{IN-Carlip:2003ne,IN-Lundgren:2006kt}. It also showed that the
phase structure of RN black holes in a cavity and RN-AdS black holes has
extensive similarities. The phase structure of various black brane systems in
a cavity was investigated in
\cite{IN-Lu:2010xt,IN-Wu:2011yu,IN-Lu:2012rm,IN-Lu:2013nt,IN-Zhou:2015yxa,IN-Xiao:2015bha}%
, and most of the systems were found to undergo Hawking-Page-like or van der
Waals-like phase transitions. Boson stars and hairy black holes in a cavity
were considered in
\cite{IN-Basu:2016srp,IN-Peng:2017gss,IN-Peng:2017squ,IN-Peng:2018abh}, which
showed that the phase structure of the gravity system in a cavity is
strikingly similar to that of holographic superconductors in the AdS gravity.
Moreover, the thermodynamic and critical behavior of de Sitter black holes in
a cavity were investigated in the extended phase space
\cite{IN-Simovic:2018tdy}. Recently, we studied Born-Infeld black holes
enclosed in a cavity in a canonical ensemble \cite{IN-Wang:2019kxp} and a
grand canonical ensemble \cite{IN-Liang:2019dni}, respectively, and found that
their phase structure has dissimilarities from that of Born-Infeld-AdS black holes.

However, although it is believed that a black hole does possess thermodynamic
quantities and extremely interesting phase structure, the statistical
description of the black hole microstates has not yet been fully understood.
Even though a complete quantum gravity theory is still absent, there have been
some attempts to understand microscopic structure of a black hole
\cite{IN-Strominger:1996sh,IN-Callan:1996dv,IN-Emparan:2006it}. Specifically,
the thermodynamic geometry method has led to many insights into microstructure
of a black hole. Following the pioneering work by Weinhold \cite{IN-Weinhold},
Ruppeiner \cite{IN-Ruppeiner:1995zz}\ introduced a Riemannian thermodynamic
entropy metric to describe the thermodynamic fluctuation theory and found a
systematic way to calculate the Ricci curvature scalar $R$ of the Ruppeiner
metric. Later, $R$ has been computed for various ordinary thermodynamic
systems, such as ideal quantum gases \cite{IN-Janyszek}, Ising models
\cite{IN-Janyszek:1989zz} and anyon gas \cite{IN-Mirza}. It showed that there
is a relation between the type of the interparticle interaction and the sign
of $R$: $R>0$ implies a repulsive interaction (e.g. ideal Bose gas) while
$R<0$ means an attractive interaction (e.g. ideal Fermi gas), and $R=0$
corresponds to no interaction (e.g. ideal gas). It has also been indicated
that, at a critical point, $\left\vert R\right\vert $ diverges as the
correlation volume for ordinary thermodynamic systems
\cite{IN-Ruppeiner:1995zz}.

The Ruppeiner geometry was subsequently exploited to probe the microstructure
of a black hole. Since the work of \cite{IN-Ferrara:1997tw}, thermodynamic
geometry has been studied for various black holes
\cite{IN-Aman:2005xk,IN-Sarkar:2006tg,IN-Shen:2005nu,IN-Quevedo:2008xn,IN-Banerjee:2010bx,IN-Astefanesei:2010bm,IN-Wei:2010yw,IN-Liu:2010sz,IN-Niu:2011tb,IN-Wei:2012ui,IN-Banerjee:2016nse,IN-Vetsov:2018dte,IN-Dimov:2019fxp,IN-Bhattacharya:2019awq}%
. The result of \cite{IN-Aman:2003ug} showed that a RN black hole has a
vanished $R$, suggesting it is a non-interacting system. On the other hand, a
Kerr-Newmann-AdS black can reduce to a RN black hole by making certain
thermodynamic variables approach zero. In these limits, a RN black hole was
found to acquire a nontrivial $R$ \cite{IN-Mirza:2007ev}, implying that the
phase space adopted in \cite{IN-Aman:2003ug} may be incomplete. Recently, by
choosing different thermodynamic coordinates, the authors of
\cite{IN-Xu:2019nnp} showed that a RN black hole is an interaction system
dominated by repulsive interaction. For a RN-AdS black hole, $R$ has been
calculated, and it was observed that $R$ can be both positive and negative,
and resembles the critical behavior of ordinary thermodynamic systems near a
critical point
\cite{IN-Aman:2003ug,IN-Shen:2005nu,IN-Niu:2011tb,IN-Sahay:2010tx,IN-Chaturvedi:2017vgq}%
. The thermodynamic geometry has been investigated recently in the extended
state space
\cite{IN-Zhang:2015ova,IN-Hendi:2015xya,IN-Wei:2015iwa,IN-Sahay:2016kex,IN-Dehyadegari:2016nkd,IN-Miao:2017cyt,IN-Li:2017xvi,IN-Du:2019poh,IN-Guo:2019oad,IN-Wei:2019uqg,IN-Wei:2019yvs,IN-Wei:2019ctz}%
, in which the cosmological constant is treated as a thermodynamic variable
and acts like a pressure term
\cite{IN-Kastor:2009wy,IN-Dolan:2011xt,IN-Gunasekaran:2012dq}. Inspired by the
Ruppeiner geometry, a RN-AdS black hole was proposed to be built of some
unknown micromolecules, interactions among which can be tested by $R$
\cite{IN-Wei:2015iwa}. Recently, a new scalar curvature $R$ was introduced for
a RN-AdS black hole, and it showed that there is a large difference between
the microstructure of a black hole and the Van der Waals fluid
\cite{IN-Wei:2019uqg}.

Although there have been a lot of work in progress on thermodynamic geometry
for various black holes of different theories of gravity in spacetimes with
differing asymptotics, little is known about thermodynamic geometry for a
black hole enclosed in a cavity. Unlike RN black holes, both RN-AdS black
holes and RN black holes in a cavity can be thermally stable and hence provide
an appropriate scenario to explore whether or not the thermodynamic geometry
is sensitive to the boundary condition of black holes. In addition, it was
recently proposed that the holographic dual of $T\bar{T}$ deformed
CFT$_{\text{2}}$ is a finite region of AdS$_{\text{3}}$ with the wall at
finite radial distance \cite{IN-McGough:2016lol,IN-Wang:2018jva}, which
further motivates us to investigate the properties of a black hole in a
cavity. To this end, we undertake a study of the thermodynamic geometry for a
RN black hole in a cavity. We report that, although the phase structure of a
RN-AdS black hole and a RN black hole in a cavity is analogous to the van der
Waals fluid, there are significant differences between the thermodynamic
geometry of a RN-AdS black hole and that of a RN black hole in a cavity.

The rest of this paper is organized as follows. In section \ref{Sec:PSRG}, we
first discuss the phase structure and thermodynamic geometry of a RN-AdS black
hole in a canonical ensemble. Although the thermodynamic geometry in the
thermodynamic coordinates of the charge $Q$ and potential $\Phi$ was
investigated in \cite{IN-Shen:2005nu,IN-Niu:2011tb}, we carry out the analysis
in a more through way with a broader survey of the parameter space and find
the $R>0$ region in the phase diagrams. The phase structure and thermodynamic
geometry of a RN black hole in a cavity are then studied in details, starting
with a discussion of its phase structure. In section \ref{Sec:CB}, the
critical behavior of the RN-AdS black hole and the RN black hole in a cavity
is obtained. We summarize our results with a brief discussion in section
\ref{Sec:DC}. For simplicity, we set $G=\hbar=c=k_{B}=1$ in this paper.

\section{Phase Structure and Thermodynamic Geometry}

\label{Sec:PSRG}

In this section, we study phase structure and thermodynamic geometry of RN-AdS
black holes and RN black holes in a cavity in a canonical ensemble. That said,
the temperature and charge of the system are fixed. Thermodynamic geometry
(Ruppeiner geometry) may provide a way to probe the microscopic structure of
black holes. Adopting the Ruppeiner approach \cite{IN-Ruppeiner:1995zz}, one
can define the Ruppeiner metric $g_{\mu\nu}^{R}$ for a thermodynamic system of
independent variables $x^{\mu}$ as%
\begin{equation}
g_{\mu\nu}^{R}=-\frac{\partial^{2}S\left(  x\right)  }{\partial x^{\mu
}\partial x^{\nu}},
\end{equation}
where $S$ is the entropy of the system. The Ruppeiner metric can be used to
measure the distance between two neighboring fluctuation states
\cite{IN-Ruppeiner:1995zz}. More interestingly, the Ricci scalar of the
Ruppeiner metric or the Ruppeiner invariant $R$ can shed light on some
information about the microscopic behavior of the system, such as the strength
and type of the dominated interaction between particles in the system.

\subsection{RN-AdS Black Holes}

The $4$-dimensional static charged RN-AdS black hole solution is described by%
\begin{align}
ds^{2}  &  =-f\left(  r\right)  dt^{2}+\frac{dr^{2}}{f\left(  r\right)
}+r^{2}\left(  d\theta^{2}+\sin^{2}\theta d\phi^{2}\right)  \text{,}%
\nonumber\\
A  &  =A_{t}\left(  r\right)  dt=-\frac{Q}{r}dt\text{,}
\label{eq:RN-AdS metric}%
\end{align}
where the metric function $f\left(  r\right)  $ is
\begin{equation}
f\left(  r\right)  =1-\frac{2M}{r}+\frac{Q^{2}}{r^{2}}+\frac{r^{2}}{l^{2}},
\end{equation}
and $l$ is the AdS radius. The parameters $M$ and $Q$ can be interpreted as
the black hole mass and charge, respectively. The Hawking temperature $T$ is
given by%
\begin{equation}
T=\frac{1}{4\pi r_{+}}\left(  1-\frac{Q^{2}}{r_{+}^{2}}+\frac{3r_{+}^{2}%
}{l^{2}}\right)  , \label{eq:RN-AdST}%
\end{equation}
where $r_{+}$ is the radius of the outer event horizon. Since $f\left(
r_{+}\right)  =0$, the mass $M$ can be expressed in terms of $r_{+}$:%
\begin{equation}
M=\frac{r_{+}}{2}\left(  1+\frac{Q^{2}}{r_{+}^{2}}+\frac{r_{+}^{2}}{l^{2}%
}\right)  . \label{eq:RN-AdS mass}%
\end{equation}
It can show that the RN-AdS black satisfies the first law of thermodynamics
\begin{equation}
dM=TdS+\Phi dQ,
\end{equation}
where $S=\pi r_{+}^{2}$ and $\Phi=Q/r_{+}$ is the entropy and potential of the
black hole, respectively. To study the phase structure of the black hole in a
canonical ensemble, we need to consider the free energy. The free energy $F$
can be obtained by computing the Euclidean action in the semiclassical
approximation and is given by%
\begin{equation}
F=M-TS.
\end{equation}

\begin{figure}[tb]
\begin{center}
\includegraphics[width=0.8\textwidth]{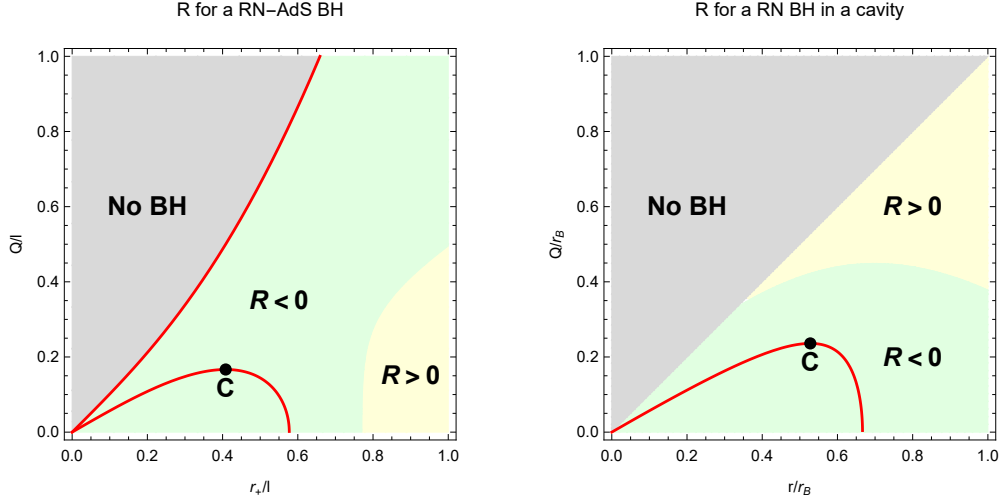}
\end{center}
\caption{{\footnotesize The Ruppeiner invariant $R$ as a function of the outer
horizon radius $r_{+}$ and the charge $Q$. \textbf{Left Panel:} For a RN-AdS
black hole, where $l$ is the AdS radius. \textbf{Right Panel:} For a RN black
hole in a cavity, where is $r_{B}$ the radius of the cavity. In the green
(yellow) regions, $R<0$ $\left(  R>0\right)  $. The red lines correspond to
$R=-\infty$. The black dots denote the critical points, which lie on the
$R=-\infty$ lines. Black hole solutions do not exist in the gray regions.}}%
\label{fig:RvrQ}%
\end{figure}

It was observed that the charge $Q$ and potential $\Phi$ of a RN-AdS black
hole play similar roles as the pressure $P$ and volume $V$ of the van der
Waals-Maxwell fluid in terms of determining the phase structure
\cite{IN-Shen:2005nu,IN-Niu:2011tb}. The correspondence $\left(
\Phi,Q\right)  \rightarrow\left(  V,P\right)  $ can establish the phase
structure of the RN-AdS black hole. In \cite{IN-Shen:2005nu}, it was also
suggested that the appropriate internal energy $U$ of the RN-AdS black hole is
given by
\begin{equation}
U=M-Q\Phi\text{,}%
\end{equation}
where the contribution of the static electricity to the black hole mass $M$ is
excluded. By analogy with the van der Waals fluid, we consider the parameter
space coordinates $x^{\mu}=\left(  U,\Phi\right)  $. Therefore, the Ruppeiner
metric becomes
\begin{equation}
g_{\mu\nu}^{R}dx^{\mu}dx^{\nu}=dUd\left(  \frac{1}{T}\right)  -\frac{Q}{T^{2}%
}d\Phi dT-\frac{1}{T}dQd\Phi.
\end{equation}
Using eqns. $\left(  \ref{eq:RN-AdST}\right)  $ and $\left(
\ref{eq:RN-AdS mass}\right)  $, we find that the expression of the Ruppeiner
invariant $R$ in terms of the horizon radius $r_{+}$ and the charge $Q$ is%
\begin{equation}
R=-\frac{\left(  Q^{2}-r_{+}^{2}\right)  ^{2}+3r_{+}^{2}\left(  10Q^{4}%
-9Q^{2}r_{+}^{2}+3r_{+}^{4}\right)  /l^{2}+18r_{+}^{6}\left(  3Q^{2}-r_{+}%
^{2}\right)  /l^{4}}{\pi\left(  r_{+}^{2}-Q^{2}+3r_{+}^{4}/l^{2}\right)
\left(  3Q^{2}-r_{+}^{2}+3r_{+}^{4}/l^{2}\right)  ^{2}}.\label{eq:RAdS}%
\end{equation}
We plot $R$ against $r_{+}$ and $Q$ in the left panel of FIG. \ref{fig:RvrQ},
where $R>0$ and $R<0$ in the yellow and green regions, respectively. The
RN-AdS black hole solution in the gray region has negative temperature and
hence is discarded. On the red lines, one has $R=-\infty$. The red line
separating the \textquotedblleft No BH\textquotedblright\ region and the
\textquotedblleft$R<0$\textquotedblright\ region is determined by $T=0$, which
shows that $R=-\infty$ for extremal RN-AdS black holes. The red line in the
green region is given by $C_{Q}^{-1}=0$, where $C_{Q}$ is the heat capacity at
constant $Q$:
\begin{equation}
C_{Q}=T\frac{\partial S}{\partial T}|_{Q}=\frac{2\pi r_{+}T}{\partial
T/\partial r_{+}|_{Q}}=\frac{8\pi^{2}r_{+}^{5}T}{3Q^{2}-r_{+}^{2}+3r_{+}%
^{4}/l^{2}}.
\end{equation}
The divergence of $C_{Q}$ usually means that the black hole would undergo a
phase transition. Moreover, using the mass representation of Ruppeiner metric,
the authors of
\cite{PSRG-Mansoori:2013pna,PSRG-Mansoori:2014oia,IN-HosseiniMansoori:2019jcs}
showed that there is a one to one correspondence between singularities of the
Ruppeiner invariant and phase transitions of $C_{Q}$. For simplicity, we
hereafter set $l=1$.

\begin{figure}[tb]
\begin{center}
\includegraphics[width=0.8\textwidth]{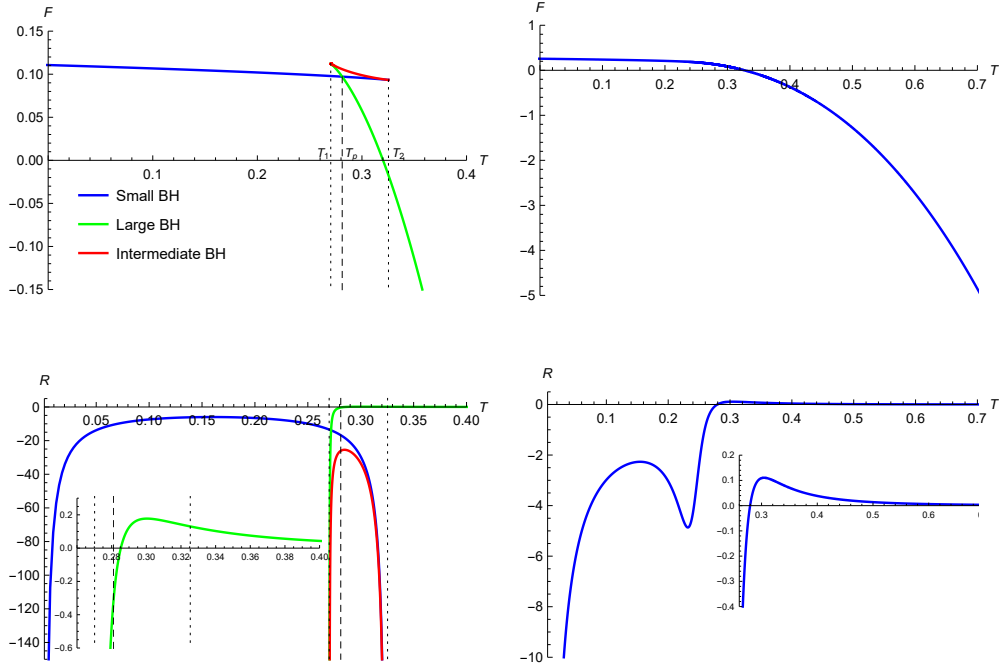}
\end{center}
\caption{{\footnotesize The free energy $F$ and the Ruppeiner invariant $R$ as
functions of the temperature $T$ for a RN-AdS black hole with a fixed charge
$Q$. \textbf{Left Column: }$Q=0.11<Q_{c}$. There are three the black hole
solutions when $T_{1}<T<T_{2}$. A LBH/SBH first-order phase transition occurs
at $T=T_{p}$, and $R=-\infty$ at $T=T_{1}$ and $T_{2}$. The inset shows that
$R>0$ for the Large BH solution at a high enough temperature. \textbf{Right
Column: }$Q=0.25>Q_{c}$. There is only one black hole solution. The inset
shows that $R>0$ when the temperature is high enough.}}%
\label{fig:TvFR}%
\end{figure}

To study the Ruppeiner invariant $R$ as a function of the temperature $T$ and
the charge $Q$, we need to use eqn. $\left(  \ref{eq:RN-AdST}\right)  $ to
express the horizon radius $r_{+}$ in terms of the temperature $T$:
$r_{+}=r_{+}(T)$. If $r_{+}(T)$ is multivalued, there is more than one black
hole solution for fixed values of $Q$ and $T$, corresponding to multiple
phases in the canonical ensemble. The critical point is an inflection point
and obtained by%
\begin{equation}
\frac{\partial T}{\partial r_{+}}=0\text{ and }\frac{\partial^{2}T}{\partial
r_{+}^{2}}=0\text{ (or equivalently }\frac{\partial Q}{\partial\Phi}=0\text{
and }\frac{\partial^{2}Q}{\partial\Phi^{2}}=0\text{),} \label{eq:cAdS}%
\end{equation}
which gives the corresponding quantities evaluated at the critical point%
\begin{equation}
\left(  r_{+c},T_{c},Q_{c},\Phi_{c}\right)  \approx\left(
0.408,0.260,0.167,0.408\right)  .
\end{equation}
When $Q<Q_{c}\approx0.167$, three black hole solutions coexist for some range
of $T$. We plot $F$ and $R$ against $T$ for $Q=0.11$ in the left column of
FIG. \ref{fig:TvFR}, which shows that there are three black hole solutions,
dubbed as Large BH, Small BH and Intermediate BH, when $T_{1}<T<T_{2}$. Note
that since $\partial^{2}F/\partial^{2}T=-C_{Q}$, the thermally stable/unstable
black hole solution has a concave downward/upward $T$-$F$ curve. So the
upper-left panel of FIG. \ref{fig:TvFR} shows that Large BH and Small BH are
thermally stable while Intermediate BH is thermally unstable. It also displays
that there is a first-order phase transition between Small BH and Large BH
occurring at $T=T_{p}$ with $T_{1}<T_{p}<T_{2}$. So the globally stable phase
is Large BH when $T>T_{p}$ and Small BH when $T<T_{p}$. As shown in the
lower-left panel of FIG. \ref{fig:TvFR}, $R$ of Small BH and Intermediate BH
is always negative, which means attractive interactions between the possible
BH molecules. For Small BH, $R=-\infty$ at $T=0$ and $T=T_{2}$, where
$\left\vert C_{Q}\right\vert =\infty$. However for Large BH, $R$ can be
negative or positive depending on the value of $T$. The inset shows that $R>0$
at a high enough temperature, and hence the interactions between the BH
molecules become repulsive. It shows that $R$ of Large BH is negative infinity
at $T=T_{1}$, where $\left\vert C_{Q}\right\vert =\infty$. Considering the
globally stable phase, one has $R=-\infty$ only for the extremal black hole
since Small BH at $T=T_{2}$ or Large BH at $T=T_{1}$ is not globally stable.
There is a crossing of $R$ of Large BH and Small BH between $T_{1}$ and
$T_{2}$. Such $R$-crossing was proposed to indicate a first-order phase
transition due to the equality of the correlation lengths for the phases at
the phase transition \cite{IN-Chaturvedi:2014vpa}. It is interesting to note
that the temperature of the $R$-crossing is different from $T_{p}$. When
$Q>Q_{c}$, there is only one black hole solution. We display $F$ and $R$
against $T$ for $Q=0.25$ in the right column of FIG. \ref{fig:TvFR}. The
upper-right panel of FIG. \ref{fig:TvFR} shows that the black hole solution is
always thermally stable. The lower-right panel of FIG. \ref{fig:TvFR} shows
that $R=-\infty$ at $T=0$, and $R>0$ when the temperature is high enough.

\begin{figure}[tb]
\begin{center}
\includegraphics[width=0.4\textwidth]{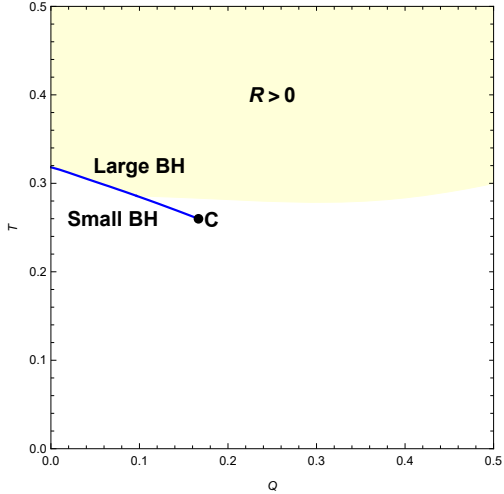}
\end{center}
\caption{{\footnotesize Phase structure of a RN-AdS black hole in the $Q$-$T$
diagram. The first-order phase transition line separating Large BH and Small
BH is displayed by the blue line and terminates at the critical point, marked
by the black dot. In the yellow region, $R>0$.}}%
\label{fig:TvQ}%
\end{figure}

In FIG. \ref{fig:TvQ}, the globally stable phase of a RN-AdS black hole, which
is the black hole solution with the minimum free energy, is displayed in the
$Q$-$T$ space. There is a LBH/SBH first-order transition line for $Q<Q_{c}$,
which terminates at the critical point. The Ruppeiner invariant $R$ is
positive in the yellow region and negative infinity at $T=0$ and the critical
point. Note that $R$ is always positive for a large enough value of $T$.
Across the first-order transition line, the microstructure of the black hole
changes while the type of the interaction among the microstructure is the same
when $0.097\lesssim Q<Q_{c}$. However, the type of the interaction changes
across the transition line when $Q\lesssim0.097$.

\begin{figure}[tb]
\begin{center}
\includegraphics[width=0.8\textwidth]{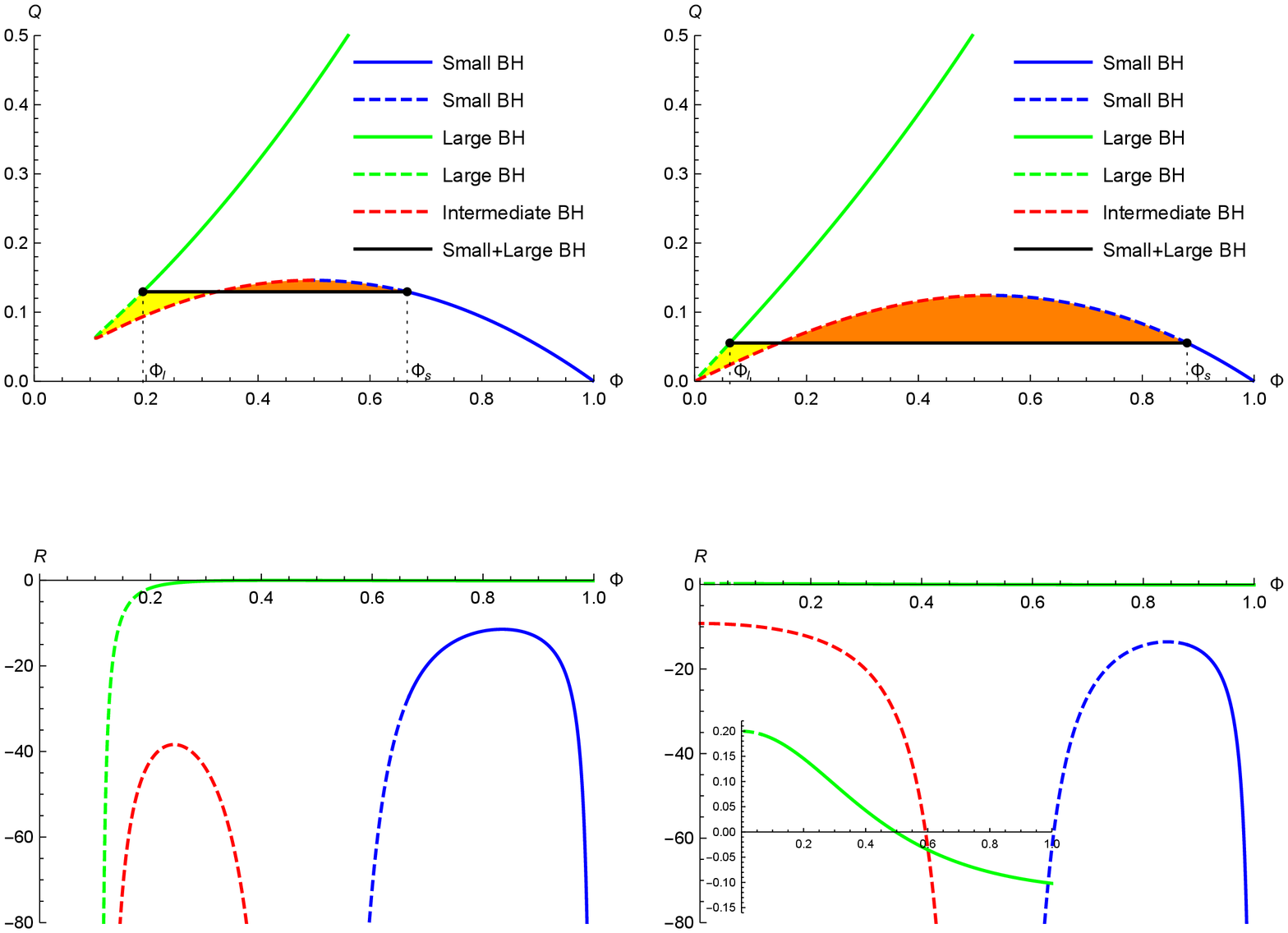}
\end{center}
\caption{{\footnotesize The charge $Q$ and the Ruppeiner invariant $R$ as
functions of the potential $\Phi$ for a RN-AdS black hole with a temperature
$T>T_{c}$, which possesses three solutions for some range of $\Phi$. The
$\Phi$-$Q$ diagrams bear resemblance to that of the Van der Waals fluid. The
coexisting lines of Large BH and Small BH are represented by the black lines,
and $\Phi_{l}/\Phi_{s}$ denotes the potential of the saturated Small/Large BH.
The blue/green dashed lines represent metastable overcharged Small
BH/undercharged Large BH while the red dashed lines denote unstable spinodal
Intermediate BH. \textbf{Left Column: }$T=0.274$. The Maxwell equal area rule
applies to determine the coexisting line, and $R$ is always negative.
\textbf{Right Column: }$T=0.3$. The Maxwell equal area rule does not apply to
determine the coexisting line due to the discontinuity of the free energy at
the origin. The inset shows that $R>0$ for the Large BH solution with a small
enough potential.}}%
\label{fig:PhivFR}%
\end{figure}

As discussed before, there is a correspondence $\left(  \Phi,Q\right)
\rightarrow\left(  V,P\right)  $ for a RN-AdS and the van der Waals fluid when
the phase structure is considered. To explore the resemblance between
thermodynamics of these two systems, we investigate $Q$ and $R$ as functions
of $T$ and $\Phi$. According to FIG. \ref{fig:TvQ}, there is no phase
coexistence when $T>T_{c}$ or $T<T_{0}\approx0.318$, where $T_{0}$ is the
temperature of the first-order phase transition line at $Q=0$. For
$T_{c}<T<T_{0}$, Large BH and Small BH can coexist for some range of $\Phi$
with some value of $Q$, which is determined by the first-order phase
transition line (i.e., $F\left(  \text{LBH}\right)  =F\left(  \text{SBH}%
\right)  $). We plot $Q$ and $R$ versus $\Phi$ for $T=0.274$ in the left
column of FIG. \ref{fig:PhivFR}, where the black line is the coexisting line
of Large BH and Small BH. Since eqn. $\left(  \ref{eq:RAdS}\right)  $ is not
applicable on the coexisting line, we do not plot $R$ for the coexisting line.
The lower-left panel of FIG. \ref{fig:PhivFR} shows that $R$ of Large BH is
always negative and becomes divergent at the minimum value of $\Phi$, where
$C_{Q}^{-1}=0$ and LBH is unstable and undercharged. For Small BH, $R$ is
negative and becomes divergent as $\Phi\rightarrow1$. The unstable and
overcharged Small BH also has $R=-\infty$ when $C_{Q}^{-1}=0$. The green and
blue solid lines represent the globally stable phase, $R$ of which is always
negative and becomes divergent as $\Phi\rightarrow1.$We plot $Q$ and $R$
versus $\Phi$ for $T=0.3$ in the right column of FIG. \ref{fig:PhivFR}. The
Maxwell equal area rule is not applicable here since there is a jump of the
free energy at $\left(  \Phi,Q\right)  =\left(  0,0\right)  $ on the
isothermal $\Phi$-$Q$ line. When $\Phi$ is small enough, the inset shows that
$R$ of Large BH becomes positive. For the globally stable phase, $R>0$ for a
small enough $\Phi$ and $R\rightarrow-\infty$ as $\Phi\rightarrow1$. Note that
$\Phi\rightarrow1$ corresponds to $M\rightarrow r_{+}\left(  1-2\pi
r_{+}T\right)  $ and $Q\rightarrow r_{+}\left(  1-2\pi r_{+}T\right)  $ with
$r_{+}\rightarrow0$.

\begin{figure}[tb]
\begin{center}
\includegraphics[width=0.8\textwidth]{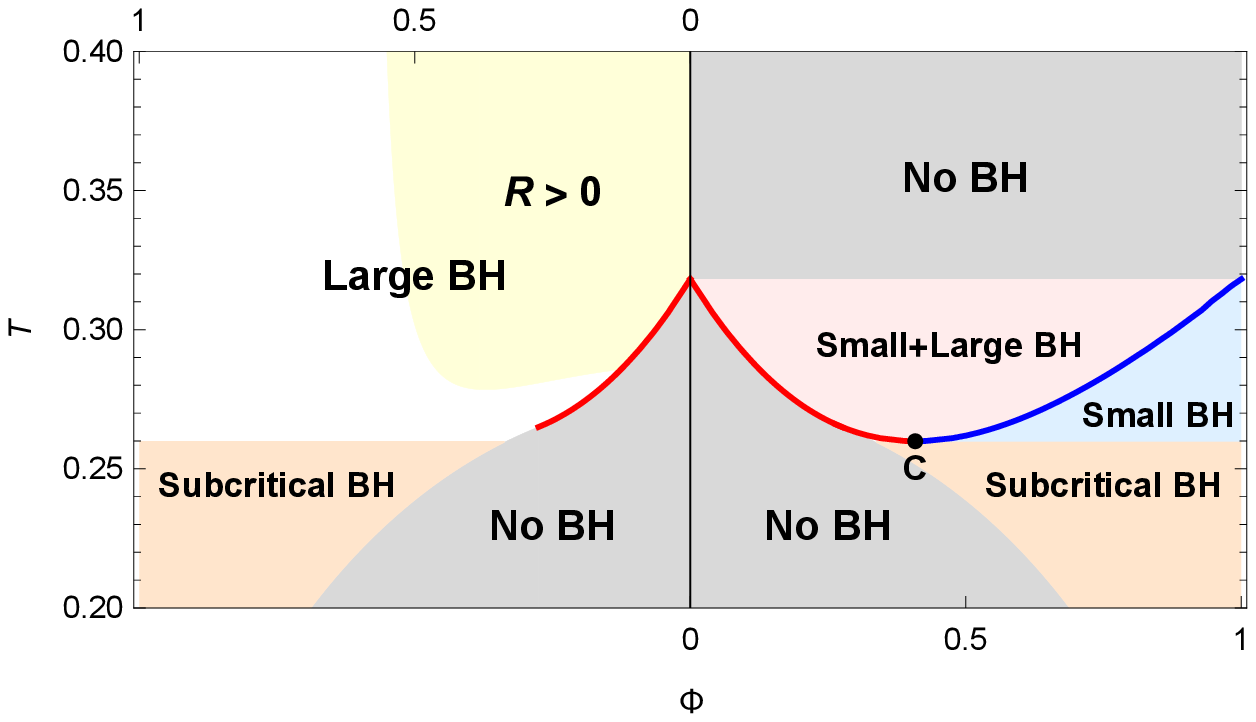}
\end{center}
\caption{{\footnotesize Phase structure of a RN-AdS black hole in the $\Phi
$-$T$ diagram. The critical point (black dot) separates the saturated Small BH
and Large BH, which are represented by the blue and red lines, respectively.
Note that there are two black hole solutions for a fixed value of $\Phi$. The
left/right frame shows the large/small $Q$ black hole solution. In the yellow
region, $R>0$. The pink region is the coexistence region of Small BH and Large
BH.}}%
\label{fig:TvPhi}%
\end{figure}

The globally stable phase of a RN-AdS black hole is plotted in the $\Phi$-$T$
space in FIG. \ref{fig:TvPhi}. It is shown in FIG. \ref{fig:PhivFR} that,
unlike the isothermal $V$-$P$ lines in the van der Waals fluid, the isothermal
$\Phi$-$Q$ lines in the RN-AdS black hole are multivalued. In fact, there are
two black hole solutions with different charges for fixed values of $\Phi$ and
$T$. The left frame of FIG. \ref{fig:TvPhi} displays the large charge solution
while the right one displays the small charge solution. Blow $T=T_{c}$, there
is no phase coexistence, and only one globally stable phase, dubbed as
Subcritical BH, exists. For Large BH, $R$ can be positive when $\Phi$ is
small. As $\Phi\rightarrow1$, $R$ of Small BH is negative infinity. When
$T\lesssim0.286$, the interactions among the microstructure of the saturated
Large BH and Small BH on the coexisting line are both attractive. However when
$T\gtrsim0.286$, the interactions of the saturated Large BH and Small BH are
repulsive and attractive, respectively.

\subsection{RN Black Holes in a Cavity}

The $4$-dimensional RN black hole solution is%
\begin{align}
ds^{2}  &  =-f\left(  r\right)  dt^{2}+\frac{dr^{2}}{f\left(  r\right)
}+r^{2}\left(  d\theta^{2}+\sin^{2}\theta d\phi^{2}\right)  \text{,}%
\nonumber\\
f\left(  r\right)   &  =1-\frac{r_{+}}{r}+\frac{Q_{b}^{2}}{r^{2}}\left(
1-\frac{r}{r_{+}}\right)  \text{, }A=A_{t}\left(  r\right)  dt=-\frac{Q_{b}%
}{r}dt\text{,}%
\end{align}
where $Q_{b}$ is the black hole charge, and $r_{+}$ is the radius of the outer
event horizon. The Hawking temperature $T_{b}$ of the RN black hole is given
by%
\begin{equation}
T_{b}=\frac{1}{4\pi r_{+}}\left(  1-\frac{Q_{b}^{2}}{r_{+}^{2}}\right)  .
\end{equation}
We now consider a thermodynamic system with a RN black holes enclosed in a
cavity. Suppose that the wall of the cavity enclosing the RN black hole is at
$r=r_{B}$, and the wall is maintained at a temperature of $T$ and a charge of
$Q$. For this system, the free energy $F$ and the thermal energy $E$ were
given in \cite{IN-Carlip:2003ne}
\begin{align}
F  &  =r_{B}\left[  1-\sqrt{f\left(  r_{B}\right)  }\right]  -\pi Tr_{+}%
^{2}\text{,}\nonumber\\
E  &  =r_{B}\left[  1-\sqrt{f\left(  r_{B}\right)  }\right]  .
\end{align}
It also showed in \cite{IN-Braden:1990hw} that the system temperature $T$ and
charge $Q$ can be related to the black hole temperature $T_{b}$ and charge
$Q_{b}$ as%
\begin{align}
Q  &  =Q_{b}\text{,}\nonumber\\
T  &  =\frac{T_{b}}{\sqrt{f\left(  r_{B}\right)  }}\text{,}%
\end{align}
which means that $T,$ measured at $r=r_{B}$, is blueshifted from $T_{b}$,
measured at $r=\infty$. The potential $\Phi$ measured on the wall is
\cite{IN-Braden:1990hw}
\begin{equation}
\Phi=\frac{A_{t}\left(  r_{B}\right)  -A_{t}\left(  r_{+}\right)  }%
{\sqrt{f\left(  r_{b}\right)  }}.
\end{equation}
As in the RN-AdS black hole, we define the internal energy $U$ by excluding
the contribution of the static electricity from the thermal energy $E$%
\begin{equation}
U=E-Q\Phi\text{.}%
\end{equation}
The physical space of $r_{+}$ is bounded by%
\begin{equation}
r_{e}\leq r_{+}\leq r_{B}\text{,} \label{eq:rBound}%
\end{equation}
where $r_{e}=Q$ is the horizon radius of the extremal black hole.

In the parameter space coordinates $x^{\mu}=\left(  U,\Phi\right)  $, the
Ruppeiner metric is
\begin{equation}
g_{\mu\nu}^{R}dx^{\mu}dx^{\nu}=dUd\left(  \frac{1}{T}\right)  -\frac{Q}{T^{2}%
}d\Phi dT-\frac{1}{T}dQd\Phi,
\end{equation}
and the Ruppeiner invariant $R$ as a function of the horizon radius $r_{+}$
and the charge $Q$ is%
\begin{equation}
R=-\frac{\left(  r_{+}^{2}-Q^{2}\right)  (r_{+}-r_{B})\left(  Q^{2}-r_{+}%
r_{B}\right)  \left[  Q^{4}(12r_{+}-5r_{B})-8Q^{2}r_{+}^{2}r_{B}+r_{+}%
^{3}r_{B}(4r_{B}-3r_{+})\right]  }{\pi r_{+}^{2}\left[  Q^{4}(6r_{+}%
-5r_{B})-2Q^{2}r_{+}\left(  r_{+}^{2}+3r_{+}r_{B}-3r_{B}^{2}\right)
+r_{+}^{3}r_{B}(3r_{+}-2r_{B})\right]  ^{2}}. \label{eq:CavityR}%
\end{equation}
We plot $R$ against $r_{+}$ and $Q$ in the right panel of FIG. \ref{fig:RvrQ},
where $R>0$ and $R<0$ in the yellow and green regions, respectively, and the
solution in the gray region does not satisfy the constraint $\left(
\ref{eq:rBound}\right)  $. The red line, on which $R=-\infty$, is also
determined by $C_{Q}^{-1}=0$, where $C_{Q}$ is the heat capacity at constant
$Q$:
\begin{equation}
C_{Q}=\frac{16\pi^{2}Tr_{+}^{9/2}(r_{B}-r_{+})^{3/2}\left(  r_{+}r_{B}%
-Q^{2}\right)  ^{3/2}}{r_{B}\left[  Q^{4}(6r_{+}-5r_{B})-2Q^{2}r_{+}\left(
r_{+}^{2}+3r_{+}r_{B}-3r_{B}^{2}\right)  +r_{+}^{3}r_{B}(3r_{+}-2r_{B}%
)\right]  }.
\end{equation}
Unlike the RN-AdS black hole, eqn. $\left(  \ref{eq:CavityR}\right)  $ gives
$R=0$ for the extremal RN black hole with $Q=r_{+}$. FIG. \ref{fig:RvrQ} shows
that the behavior of $R$ of RN-AdS black holes is quite different from that of
RN black holes in a cavity. For simplicity, we hereafter set $r_{B}=1$.

\begin{figure}[tb]
\begin{center}
\includegraphics[width=0.8\textwidth]{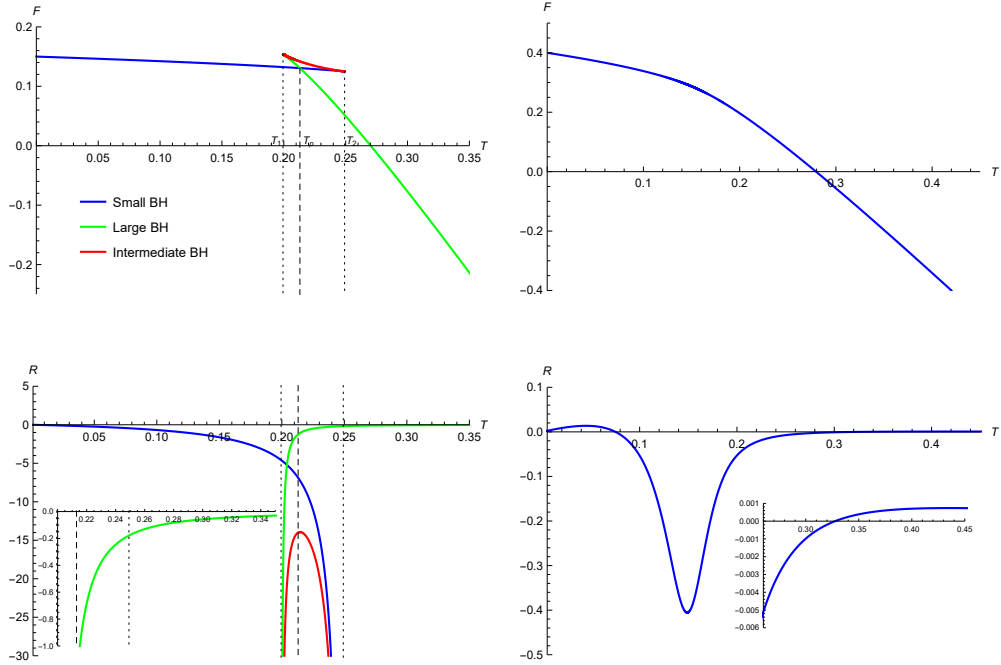}
\end{center}
\caption{{\footnotesize The free energy $F$ and the Ruppeiner invariant $R$ as
functions of the temperature $T$ for a RN black hole in a cavity with a fixed
charge $Q$. \textbf{Left Column: }$Q=0.15<Q_{c}$. Three black hole solutions
coexist for $T_{1}<T<T_{2}$. A LBH/SBH first-order phase transition occurs at
$T=T_{p}$, and $R=-\infty$ at $T=T_{1}$ and $T_{2}$. The $T$-$R$ diagram and
the inset show that $R$ is always negative. \textbf{Right Column:
}$Q=0.4>Q_{c}$. There is only one black hole solution. As shown in the $T$-$R$
diagram and the inset, $R>0$ when the temperature is high enough or low
enough.}}%
\label{fig:CTvFR}%
\end{figure}

It was observed that the phase structure of a RN black hole in a cavity\ is
strikingly similar to that of a RN-AdS black hole in the $Q$-$T$ space of a
canonical ensemble \cite{IN-Carlip:2003ne,IN-Lundgren:2006kt}. Actually, a van
der Waals-like phase transition occurs in both cases. Similar to the RN-AdS
black hole, the RN black hole in a cavity possesses a critical point, which is
determined by%
\begin{equation}
\frac{\partial T}{\partial r_{+}}=0\text{ and }\frac{\partial^{2}T}{\partial
r_{+}^{2}}=0\text{.}%
\end{equation}
Solving the above equations gives quantities evaluated at the critical point%
\begin{equation}
\left(  r_{+c},T_{c},Q_{c},\Phi_{c}\right)  \approx\left(
0.528,0.186,0.236,0.325\right)  .
\end{equation}
The number of the solution(s) $r_{+}(T)$ to $T=T\left(  r_{+}\right)  $
depends on the value of $Q$. When $Q<Q_{c}\approx0.236$, there are three
solutions coexisting for some range of $T$. We plot $F$ and $R$ against $T$
for $Q=0.15$ in the left column of FIG. \ref{fig:CTvFR}, which shows that
three solutions, dubbed as Large BH, Small BH and Intermediate BH, coexist
when $T_{1}<T<T_{2}$. Note that Large BH and Small BH are the thermally stable
while Intermediate BH is thermally unstable. The system undergoes a
first-order phase transition between Small BH and Large BH occurring at
$T=T_{p}$ with $T_{1}<T_{p}<T_{2}$. The lower-left panel of FIG.
\ref{fig:CTvFR} and the inset display that $R$ of all three solutions is
always negative, which means attractive interactions between the possible BH
molecules. There is also a crossing of $R$ of Large BH and Small BH between
$T_{1}$ and $T_{2}$. For Large BH/Small BH, $R=-\infty$ at $T=T_{1}/T_{2}$,
where $\left\vert C_{Q}\right\vert =\infty$. However, $R=0$ when $T=0$, which
gives that $R$ of the globally stable phase is always finite. When $Q>Q_{c}$,
there is only one solution. We depict $F$ and $R$ as functions of $T$ for
$Q=0.4$ in the right column of FIG. \ref{fig:CTvFR}. The upper-right panel of
FIG. \ref{fig:CTvFR} shows that the solution is always thermally stable. The
lower-right panel of FIG. \ref{fig:CTvFR} and the inset show that $R$ is
always finite, and $R>0$ when the temperature is high enough or low enough.

\begin{figure}[tb]
\begin{center}
\includegraphics[width=0.4\textwidth]{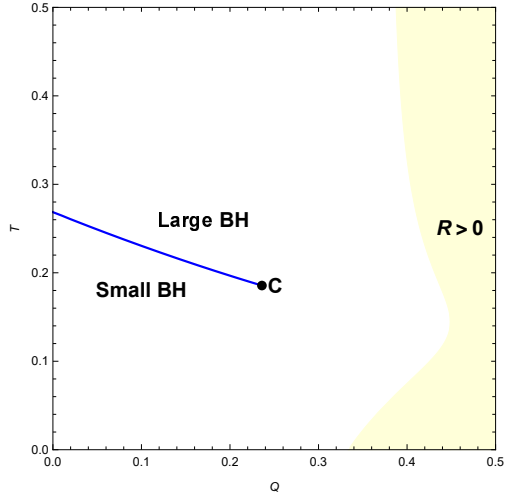}
\end{center}
\caption{{\footnotesize Phase structure of a RN black hole in a cavity in the
$Q$-$T$ diagram. The first-order phase transition line separating Large BH and
Small BH is displayed by the blue line and terminates at the critical points
(black dot). In the yellow region, $R>0$.}}%
\label{fig:CTvQ}%
\end{figure}

In FIG. \ref{fig:CTvQ}, the globally stable phase of a RN black hole in a
cavity is displayed in the $Q$-$T$ space. There is a LBH/SBH first-order
transition line for $Q<Q_{c}$, which terminates at the critical point. The
Ruppeiner invariant $R$ is positive in the yellow region and negative infinity
only at the critical point. Note that $R$ is always positive when $Q$ is large
enough. Since $R$ of Large BH above and Small BH below the transition line is
negative, the type of the interaction among the microstructure of the black
hole stays the same when the system undergoes the first-order transition.

\begin{figure}[tb]
\begin{center}
\includegraphics[width=0.8\textwidth]{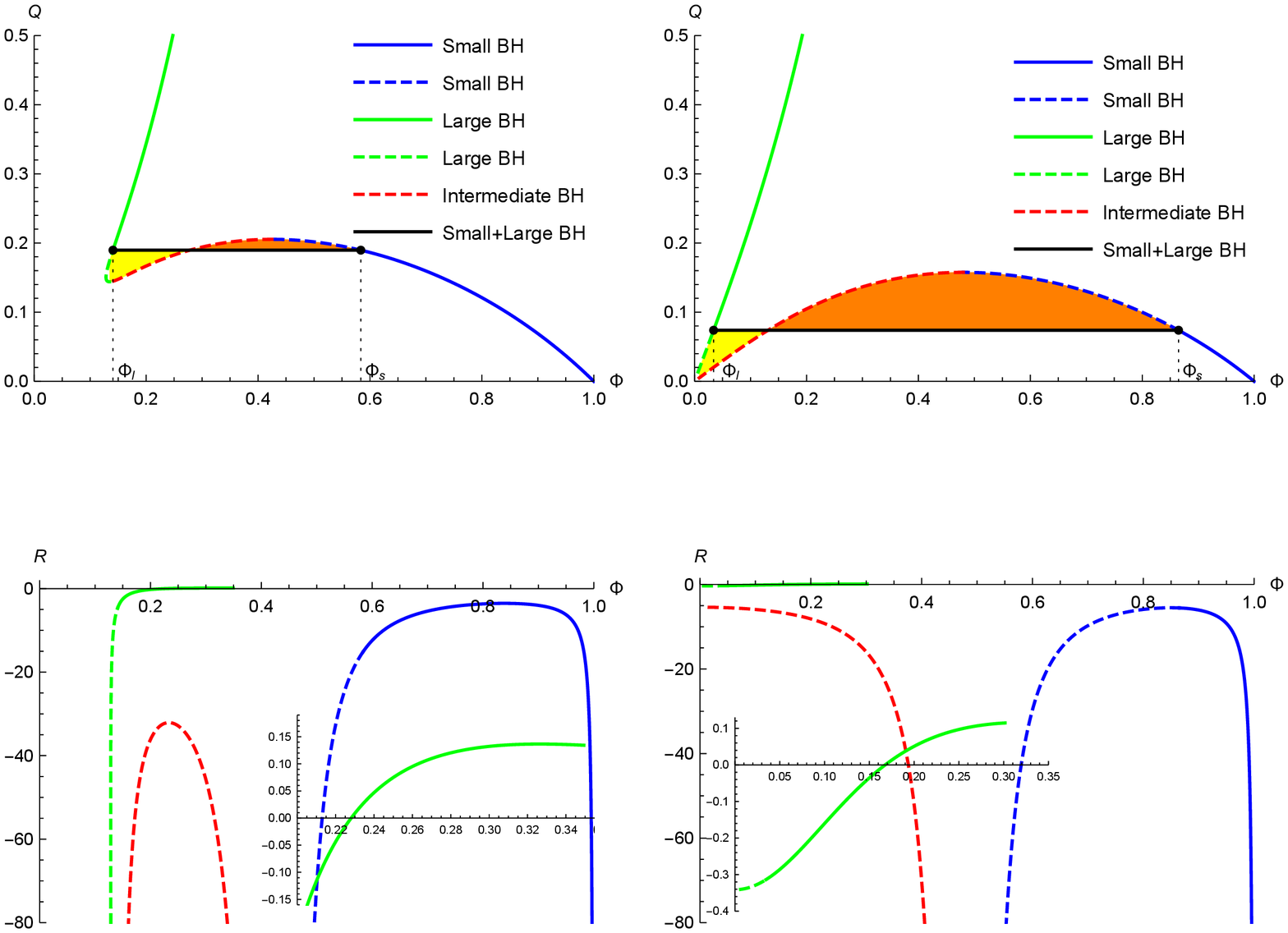}
\end{center}
\caption{{\footnotesize The charge $Q$ and the Ruppeiner invariant $R$ as
functions of the potential $\Phi$ for a RN black hole in a cavity with a
temperature $T>T_{c}$, which possesses three solutions for some range of
$\Phi$. The black lines denote the coexisting lines of Large BH and Small BH,
and $\Phi_{s}/\Phi_{l}$ denotes the potential of the saturated Small/Large BH.
The blue/green dashed lines represent metastable overcharged Small
BH/undercharged Large BH while the red dashed lines denote unstable spinodal
Intermediate BH. \textbf{Left Column: }$T=0.2$. $R$ is always negative.
\textbf{Right Column: }$T=0.24$. The inset shows that $R>0$ for the Large BH
solution with a large enough potential.}}%
\label{fig:CPhivFR}%
\end{figure}

We now investigate the phase structure of a RN black hole in a cavity in the
$\Phi$-$T$ space, which has not been discussed in the literature. Similar to a
RN-AdS black hole, there is phase coexistence of Large BH and Small BH when
$T_{c}<T<T_{0}\approx0.269$. We plot $Q$ and $R$ versus $\Phi$ for $T=0.2$ in
the left column of FIG. \ref{fig:CPhivFR}, where the black line is the
coexisting line of Large BH and Small BH. The upper bound on the $\Phi$ of
Large BH comes from $r_{+}\leq1$. The lower-left panel of FIG.
\ref{fig:CPhivFR} displays that $R$ of all three solutions is always negative
and becomes divergent when $C_{Q}^{-1}=0$ or $\Phi\rightarrow1$. Note that
$\Phi\rightarrow1$ corresponds to $Q\rightarrow r_{+}\left(  1-2\pi
r_{+}T\right)  $ with $r_{+}\rightarrow0$. We plot $Q$ and $R$ versus $\Phi$
for $T=0.3$ in the right column of FIG. \ref{fig:CPhivFR}. The inset shows
that $R$ of Large BH is always finite and becomes positive when $\Phi$ is
large enough. For Small BH, $R<0$ and $R\rightarrow-\infty$ as $\Phi
\rightarrow1$.

\begin{figure}[tb]
\begin{center}
\includegraphics[width=0.8\textwidth]{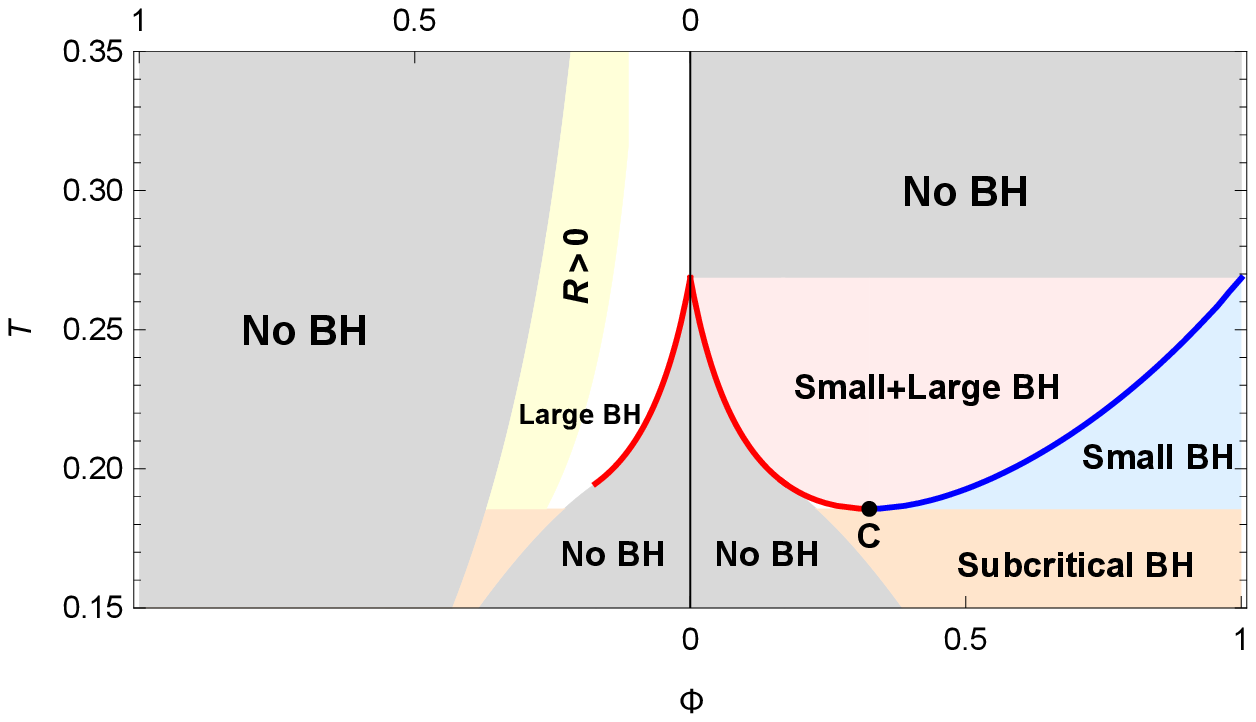}
\end{center}
\caption{{\footnotesize Phase structure of a RN black hole in a cavity in the
$\Phi$-$T$ diagram. The saturated Small BH (blue line) and Large BH (red line)
are separated by the critical point (black dot). For a fixed value of $\Phi$,
there exist two black hole solutions, which are displayed in the left and
right frames, respectively. The pink region is the coexistence region of Small
BH and Large BH, and $R>0$ in the yellow region.}}%
\label{fig:CTvPhi}%
\end{figure}

The globally stable phase of a RN-AdS black hole is plotted in the $\Phi$-$T$
space in FIG. \ref{fig:CTvPhi}, the left/right frame of which displays the
large/small charge solution. There is Subcritical BH and no phase coexistence
blow $T=T_{c}$. For Large BH, $R$ can be positive when $\Phi$ is large enough.
As $\Phi\rightarrow1$, $R$ of Small BH approaches negative infinity. Unlike
the RN-AdS black hole case, the interactions among the microstructure of the
saturated Large BH and Small BH on the coexisting line are always attractive.

\section{Critical Behavior}

\label{Sec:CB}

In this section, we investigate the thermodynamic behavior near the critical
point. The critical behavior in a RN-AdS black hole has been discussed in
\cite{IN-Shen:2005nu,IN-Niu:2011tb}, which showed that the critical exponents
for a RN-AdS black hole and the Van der Waals fluid are identical. However,
the critical exponents for a RN black hole in cavity have not been calculated yet.

First, we define%
\begin{equation}
t=\frac{T-T_{c}}{T_{c}}\text{ and }\phi=\frac{\Phi-\Phi_{c}}{\Phi_{c}}.
\end{equation}
In the neighborhood of the critical point, we can expand $Q$ in terms of $t$
and $\phi$%
\begin{equation}
Q=\sum\limits_{i,j=0}q_{ij}t^{i}\phi^{j}, \label{eq:Qexp}%
\end{equation}
where $q_{00}=Q_{c}$, and eqn. $\left(  \ref{eq:cAdS}\right)  $ gives
$q_{01}=q_{02}=0$. Specifically for a RN-AdS black hole and a RN black hole in
a cavity, we find%
\begin{align}
Q  &  =\frac{1}{6}-\frac{\phi^{3}}{3}+t\left(  -\frac{2}{3}+2\phi-8\phi
^{2}+\frac{28}{3}\phi^{3}\right)  +\cdots\text{,}\nonumber\\
Q  &  =0.236-0.182\phi^{3}+t\left(  -0.618+1.118\phi-2.834\phi^{2}%
+2.045\phi^{3}\right)  +\cdots\text{,} \label{eq:Qtphi}%
\end{align}
respectively. Near the critical point, the Maxwell equal area rule is
applicable and gives%
\begin{equation}
\int_{\Phi_{s}}^{\Phi_{l}}\Phi dQ=0\text{ with }Q\left(  \Phi_{l}\right)
=Q\left(  \Phi_{s}\right)  , \label{eq:MaxE}%
\end{equation}
where $\Phi_{s}$ and $\Phi_{l}$ denote the potential of the saturated Small BH
and Large BH, respectively. Substituting eqn. $\left(  \ref{eq:Qexp}\right)  $
into eqn. $\left(  \ref{eq:MaxE}\right)  $, we have%
\begin{equation}
\phi_{l}=-\phi_{s}\text{, }\phi_{l}=\sqrt{-\frac{q_{11}}{q_{03}}t}%
+\mathcal{O}\left(  t\right)  , \label{eq:phit}%
\end{equation}
where $\phi_{l/s}\equiv\left(  \Phi_{l/s}-\Phi_{c}\right)  /\Phi_{c}$.

The critical exponents $\alpha$, $\beta$, $\gamma$ and $\delta$ are defined as follows:

\begin{itemize}
\item Exponent $\alpha$ describes the critical behavior of the specific heat
at constant $\Phi$: $C_{\Phi}\propto\left\vert t\right\vert ^{\alpha}$. We
find that, at the critical point,
\begin{equation}
C_{\Phi}=\left\{
\begin{array}
[c]{c}%
-4.189\text{ for a RN-AdS BH,}\\
-3.167\text{ for a RN BH in a cavity,}%
\end{array}
\right.  \label{eq:Cphi}%
\end{equation}
which gives $\alpha=0$ in both cases.

\item Exponent $\beta$ describes the critical behavior of the order parameter
$\phi_{l}-\phi_{s}$: $\phi_{l}-\phi_{s}\propto\left\vert t\right\vert ^{\beta
}$. Eqn. $\left(  \ref{eq:phit}\right)  $ gives%
\[
\phi_{l}-\phi_{s}\sim\left\{
\begin{array}
[c]{c}%
4.899\sqrt{t}\text{ for a RN-AdS BH,}\\
4.952\sqrt{t}\text{ for a RN BH in a cavity,}%
\end{array}
\right.
\]
which leads to $\beta=1/2$ in both cases.

\item Exponent $\gamma$ describes the critical behavior of the isothermal
compressibility:
\begin{equation}
\kappa_{T}=-\frac{1}{\Phi}\frac{\partial\Phi}{\partial Q}|_{T}\propto
\left\vert t\right\vert ^{-\gamma}.
\end{equation}
Approaching the critical point along the coexistence curve, we use eqn.
$\left(  \ref{eq:Qexp}\right)  $ to obtain%
\begin{equation}
\kappa_{T}\sim\frac{1}{2q_{11}t}\sim\left\{
\begin{array}
[c]{c}%
0.25t^{-1}\text{ for a RN-AdS BH,}\\
0.447t^{-1}\text{ for a RN BH in a cavity,}%
\end{array}
\right.
\end{equation}
which gives $\gamma=1$ in both cases.

\item Exponent $\delta$ describes the critical behavior of $\left\vert
Q-Q_{c}\right\vert \propto\left\vert \phi\right\vert ^{\delta}$ when $T=T_{c}%
$. On the critical isotherm $T=T_{c}$, eqn. $\left(  \ref{eq:Qexp}\right)  $
reduces to%
\begin{equation}
\left\vert Q-Q_{c}\right\vert \sim\left\vert q_{03}\right\vert \phi^{3}%
\sim\left\{
\begin{array}
[c]{c}%
\phi^{3}/3\text{ for a RN-AdS BH,}\\
0.182\phi^{3}\text{ for a RN BH in a cavity,}%
\end{array}
\right.
\end{equation}
which gives $\delta=3$ in both cases.
\end{itemize}

Our results show that the critical exponents of a RN black hole in cavity are
also identical to these of the Van der Waals fluid predicted by the mean field
theory. The critical exponents are believed to be universal since they are
insensitive to the details of the physical system.

Finally, we study the critical behavior of the Ruppeiner invariant $R$ by
expanding $R$ along the saturated Small BH and Large BH curves near the
critical point. For the saturated Small BH and Large BH, we have%
\begin{equation}
r_{+l/s}-r_{+c}\simeq\frac{\Phi_{l/s}-\Phi_{c}}{\partial\Phi/\partial
r|_{Q=Q_{c}}}\simeq\pm\frac{\Phi_{c}}{\partial\Phi/\partial r|_{Q=Q_{c}}}%
\sqrt{-\frac{q_{11}}{q_{03}}t}, \label{eq:rsat}%
\end{equation}
where $+/-$ is for the saturated Large BH/Small BH, and $r_{+l/s}$ is the
horizon radius of the saturated Large BH/Small BH. In addition, the charge $Q$
for the coexisting line is given by%
\begin{equation}
Q\simeq Q_{c}+q_{03}\phi_{l/s}^{3}+q_{10}t+q_{11}t\phi_{l/s}\simeq
Q_{c}+q_{10}t. \label{eq:Qsat}%
\end{equation}
Plugging eqns. $\left(  \ref{eq:rsat}\right)  $ and $\left(  \ref{eq:Qsat}%
\right)  $ into eqns. $\left(  \ref{eq:RAdS}\right)  $ and $\left(
\ref{eq:CavityR}\right)  $, we can expand $R$ at $t=0$ and obtain%
\begin{equation}
R\sim\left\{
\begin{array}
[c]{c}%
-0.030t^{-2}\text{ for a RN-AdS BH,}\\
-0.016t^{-2}\text{ for a RN BH in a cavity,}%
\end{array}
\right.  \label{eq:Rcr}%
\end{equation}
where the leading order terms of the expansions are the same for the saturated
Small BH and Large BH. From eqns. $\left(  \ref{eq:Cphi}\right)  $ and
$\left(  \ref{eq:Rcr}\right)  $, we find%
\begin{equation}
\lim_{t\rightarrow0}RC_{\Phi}t^{2}=\left\{
\begin{array}
[c]{c}%
1/8\text{ for a RN-AdS BH,}\\
\text{ }0.051\text{ for a RN BH in a cavity,}%
\end{array}
\right.  \label{eq:RCcr}%
\end{equation}
which shows that the critical value of $RC_{\Phi}t^{2}$ may depend on the
boundary conditions of the black hole. In \cite{IN-Wei:2019uqg}, $R$ of an
RN-AdS black hole was calculated in the thermodynamic coordinates $x^{\mu
}=\left(  T,V\right)  $, and it was found $\lim_{t\rightarrow0}RC_{V}%
t^{2}=-1/8$, which agrees with the numerical result of the Van der Waals
fluid. However, there is a sign difference between our AdS result in eqn.
$\left(  \ref{eq:RCcr}\right)  $ and the result in \cite{IN-Wei:2019uqg},
which comes from $C_{\Phi}<0$ at the critical point for a RN-AdS black hole.

\section{Discussion and Conclusion}

\label{Sec:DC}

In this paper, we studied the phase structure, thermodynamic geometry and
critical behavior of RN black holes in a canonical ensemble by considering two
boundary conditions, namely the asymptotically AdS boundary and the Dirichlet
boundary in the asymptotically flat spacetime. The phase structure of a RN-AdS
black hole and a RN black hole in a cavity in the $Q$-$T$ space was displayed
in FIGs. \ref{fig:TvQ} and \ref{fig:CTvQ}, respectively, and the phase
structure in the $\Phi$-$T$ space was shown in FIGs. \ref{fig:TvPhi} and
\ref{fig:CTvPhi}. It showed that the phase structure in both cases are quite
similar. Specifically, with a correspondence $\left(  \Phi,Q\right)
\rightarrow\left(  V,P\right)  $, the phase structure of the RN-AdS black hole
and the RN black hole in a cavity was rather similar to that of the Van der
Waals fluid. We also calculated the critical exponents for the RN-AdS black
hole and the RN black hole in a cavity and found they are identical to these
of the Van der Waals fluid.

\begin{table}[ptb]
\begin{center}%
\begin{tabular}
[c]{l|c|c}\hline
Black hole solution & \thinspace$R>0$\thinspace\thinspace & $\,\,|R|$
divergence\thinspace\thinspace\\\hline
SBH in AdS & none & $T=0$ and $C_{Q}^{-1}=0$\\
SBH in Cavity & none & $C_{Q}^{-1}=0$\\
LBH in AdS & large $T$ & $C_{Q}^{-1}=0$\\
LBH in Cavity & none & $C_{Q}^{-1}=0$\\
BH in AdS & large $T$ & $T=0$\\
BH in Cavity & large $T$ or small $T$ or all $T$ & none\\
GSP in AdS & large $T$ & $T=0$ and critical point\\
GSP in Cavity & large $Q$ & critical point\\\hline
\end{tabular}
\end{center}
\caption{{\footnotesize The Ruppeiner invariant $R$ for the black hole
solutions in the AdS and cavity cases. Tabulated are the regions where $R>0$
and the possible divergences of $R$. BH denotes the single black hole solution
when $Q>Q_{c}$. GSP strands for \textquotedblleft globally stable
phase\textquotedblright.}}%
\label{tab:1}%
\end{table}

However, we found that the thermodynamic geometry in the AdS and cavity cases
is quite different. The Ruppeiner invariant $R$ as a function of the horizon
radius $r_{+}$ and the charge $Q$ was obtained in the AdS and cavity cases and
plotted in FIG. \ref{fig:RvrQ}, which showed that the $R>0$ regions (yellow
regions) are dissimilar in the two cases. Moreover, the extremal RN-AdS black
hole has $R=-\infty$ while $R=0$ for the extremal RN black hole in a cavity.
We also discussed $R$ as a function of $T$ and $Q$ and summarize the results
for the AdS and cavity cases in Table \ref{tab:1}. Moreover, we found that the
interactions among the microstructure of the black hole always stay attractive
before and after the LBH/SBH first-order transition in the cavity case.
However for a RN-AdS black hole with $Q\lesssim0.097$, the type of the
interactions changes when the black hole undergoes the phase transition. The
Ruppeiner invariant $R$ as a function of $\Phi$ and $T$ was also discussed,
and it showed that $R\rightarrow-\infty$ as $\Phi\rightarrow1$ with fixed $T$.
FIGs. \ref{fig:TvPhi} and \ref{fig:CTvPhi} displayed that Large BH with a
small enough $\Phi$ can have $R>0$ in the AdS case while Large BH with a large
enough $\Phi$ can have $R>0$ in the cavity case. Furthermore, the interactions
among the microstructure of the saturated Large BH and Small BH on the
coexisting line are always attractive in the cavity case. However in the AdS
case, the interactions of the saturated Large BH and Small BH are repulsive
and attractive, respectively, when $T\gtrsim0.286$.

In summary, although the phase structure of a RN-AdS black hole and a RN black
hole in a cavity is strikingly similar, we found that the thermodynamic
geometry in the two cases behaves rather differently. It seems that the
Ruppeiner invariant $R$ depends not only on what is inside the horizon, but
also on the imposed boundary condition. Our results show that either $R$
encodes more than the nature of black hole microscopic properties, or there
may be a connection between the black hole microstates and the boundary
condition. It would be interesting to study how $R$ depends on the boundary
condition in other black hole systems, which may shine further light on the
connection between the internal microstructure of black holes and the boundary condition.

\begin{acknowledgments}
We are grateful to S. A. Hosseini Mansoori, Shuxuan Ying and Zhipeng Zhang for
useful discussions and valuable comments. This work is supported in part by
NSFC (Grant No. 11875196, 11375121 and 11005016).
\end{acknowledgments}

\end{document}